\documentclass{kluwer}   

\usepackage{graphicx}
\newdisplay{guess}{Conjecture}

\begin{document}                                                                                   
\begin{article}
\begin{opening}         
\title{Chemical evolution of Elliptical Galaxies and the ICM} 
\author{  Antonio \surname{Pipino} and Francesca \surname{Matteucci}}  
\runningauthor{Pipino \& Matteucci}
\runningtitle{Chemical evolution of Elliptical Galaxies and the ICM}
\institute{Dipartimento di Astronomia, Universit\'{a} di Trieste, via
	G. B. Tiepolo 11, 34131 Trieste, Italy}

\begin{abstract}
We present a new model for the chemical evolution of elliptical galaxies
taking into account SN feedback, detailed nucleosynthesis and galactic winds.
We discuss the effect of galactic winds on the chemical enrichment of the ICM
and compute the energy per particle injected by the galaxies into the ICM. 
\end{abstract}
\end{opening}           

\section{The chemical evolution model for Ellipticals}  

We present the results obtained with a new chemical evolution model for ellipticals,
based on the multi-zone model by Martinelli et al. (1998), with improved SN feedback and different
prescriptions for the stellar nucleosynthesis (see Pipino et al. 2002). This model can be applied to infer the age
for a single galaxy (as MS1512-cB58, see Matteucci \& Pipino 2002), to test different
scenarios for galaxy formation (i.e. monolithic versus hierarchical)
and to predict the evolution of the chemical abundances and the energy per particle 
released by elliptical galaxies through galactic winds into the ICM.
      
 \subsection{Galactic models}  
	
We adopt two different recipes for SNIa and SNII, respectively.
For SNII we assume that the evolution of the energy in the 'snowplow'
phase is regulated by Cioffi, McKee \& Bertschinger (1988) cooling time 
(for a detailed discussion see Pipino et al. 2002).
This cooling time takes into account the effect of metallicity, 
and therefore can model in a self-consistent way the evolution of a SNR in the ISM.

On the other hand, following Recchi et al. (2001), we assume that SNIa, which explode
in a medium already heated by SNII, contribute with the total amount of their
energy budget, without radiative losses.
In this way we obtain a total (i.e. SNIa+II) mean efficiency of energy
release by SN to the ISM of $\sim$ 20\% for our best model, and derive a maximum possible efficiency 
(in order to have realistic models) of about $\sim$ 35\% .
 
We study the effects of two different sets of stellar yields on the enrichment of the ICM:
\bf a) \rm For low and intermediate mass stars ($0.8 \le M/M_{\odot} \le 8$)
the yields by Renzini and Voli (1981).
For massive stars, type II SNe ($M> 10M_{\odot}$) the yields by 
Woosley and Weaver (1995), their case B.
\bf b) \rm For low and intermediate mass stars the yields by van den Hoek \& Groenewegen (1997).
For massive stars, type II SNe, the yields by Thielemann et al. (1996)
In both cases we use the yields by Nomoto et al. (1997) for SNIa.

Here we present the results obtained with our best model: a 
multi-zone model with Salpeter IMF, and the choice \bf a) \rm for stellar yields.
The adopted cosmology is $\Omega_m=0.3$, $\Omega_{\Lambda}=0.7$
and $H_o=70 \rm \,km\,  s^{-1}\, Mpc^{-1}$ and the galaxies form at $z_f=8$.
In order to check our results we compared our prediction for the Fe enrichment
of the ICM, with predictions from models with different prescriptions (see Pipino et al. 2002).
Our models assume increasing efficiency of SF with galactic mass 
which makes larger galaxies undergo galactic wind before the smaller ones.
This implies shorter star-formation timescales for more massive galaxies and
allows us to reproduce the increasing trend 
of [Mg/Fe] with galactic mass (Matteucci 1994).

\subsection{The age of the galaxy MS1512-cB58}   
We can constrain the age of the Lyman break galaxy MS1512-cB58 (showing that its chemical properties are consistent
with those of an elliptical galaxy or a bulge in the early phases of its life, Matteucci \& Pipino 2002).
In particular, we find that this age should be $\sim 30 Myr$ by
comparing our model predictions on the chemical abundances with the observations of Pettini et al. (2002).
  \begin{figure}
           \includegraphics[width=6cm,height=6.5cm]{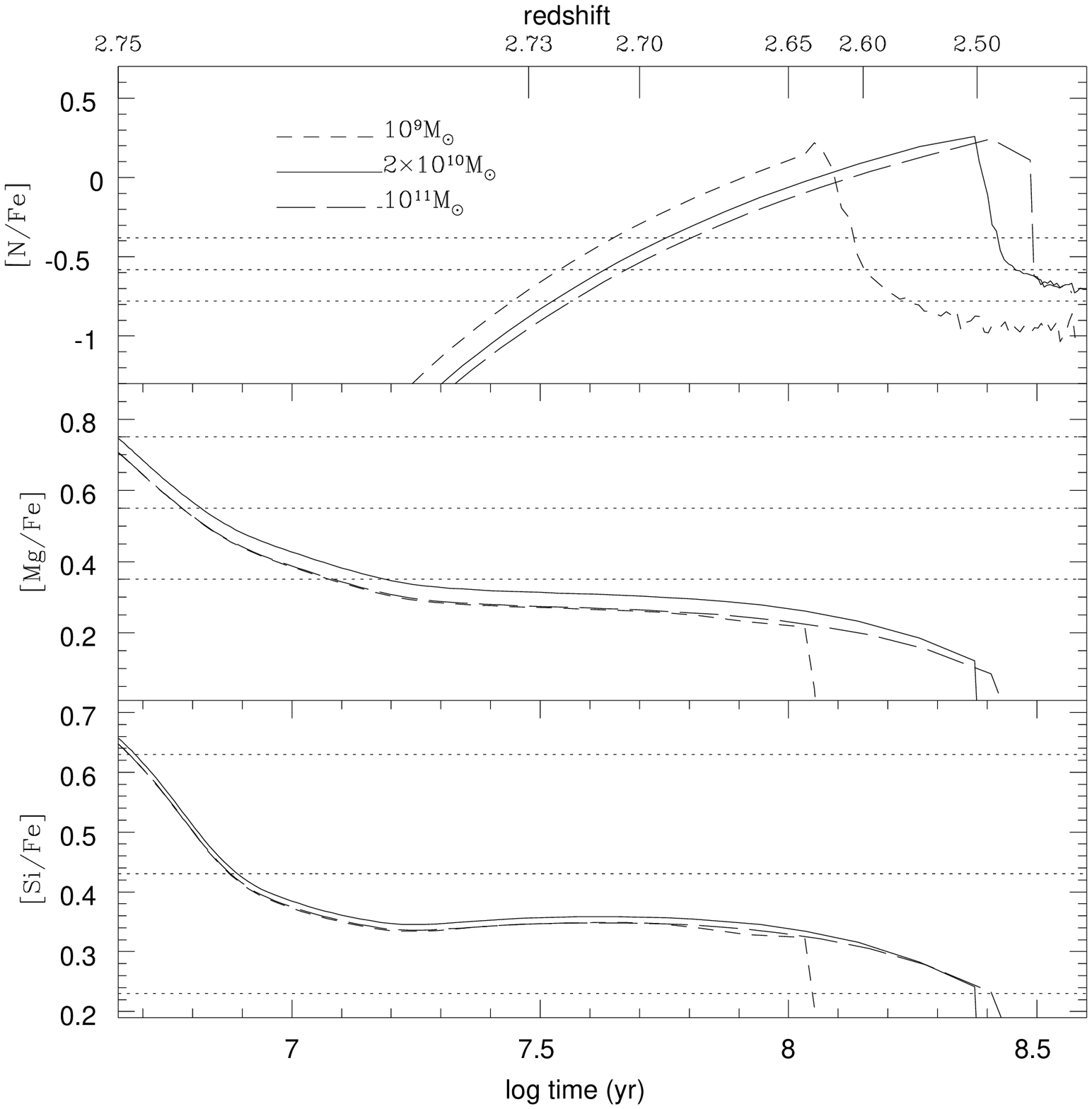}
           \includegraphics[width=6cm,height=6.5cm]{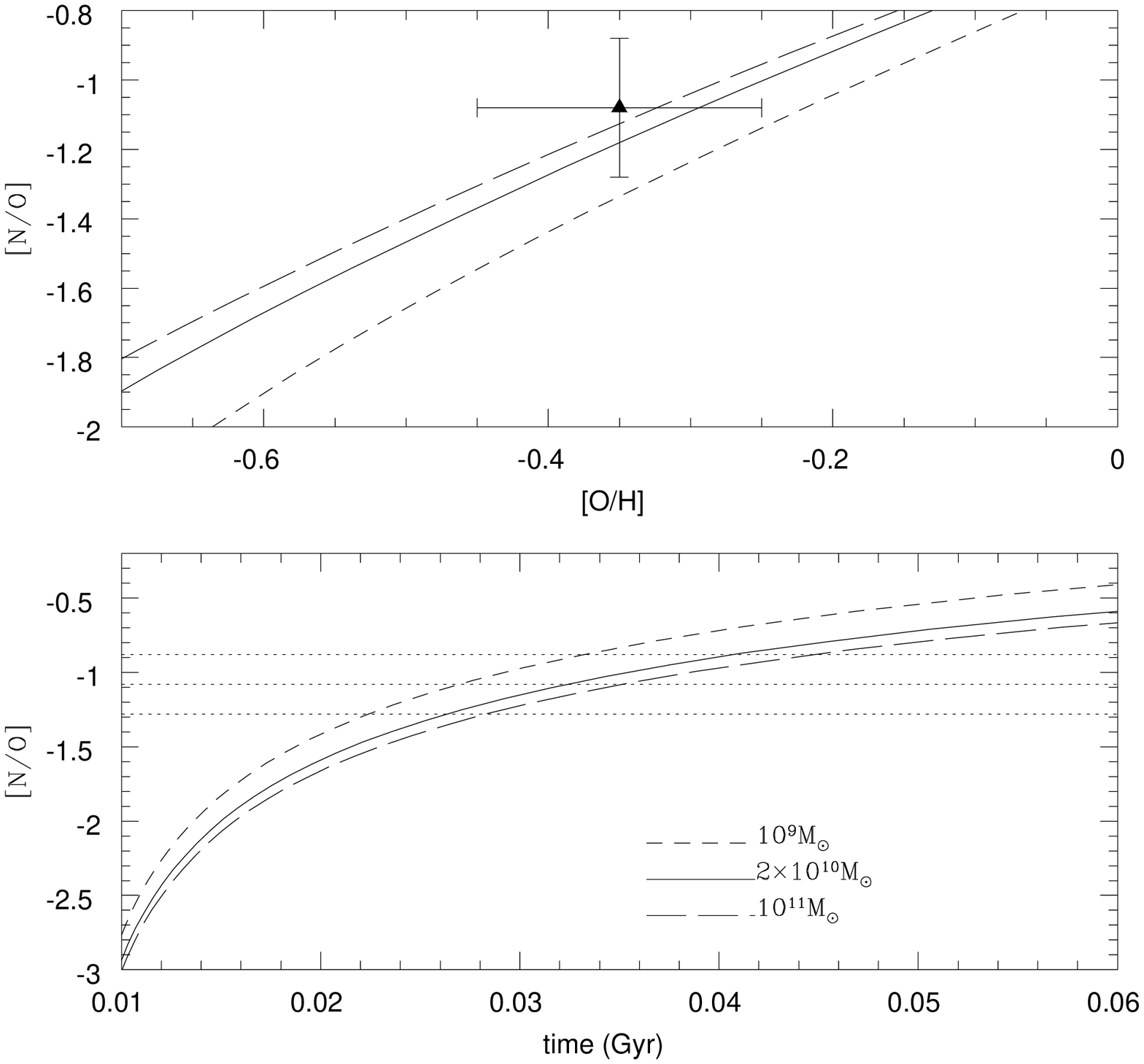}  
	   \caption{\tiny Evolution of abundances and abundance ratios in the ISM for ellipticals of different initial mass.
The observed values for MS1512-cB58 (Pettini et al. 2002) are indicated (see text for details).} 
           \end{figure}
In the first figure we show the prediction for [X/Fe] vs. time (left panel), assuming that roughly
half of the iron is hidden in dust grains, compared to the observations (showed as horizontal lines)
with their errors. 
The right panel allows us to make a stronger conclusion on the age, because neither O nor N
are dust depleted. 


       \section{Evolution of abundances and energy in the ICM}
To compute the total masses of the chemical elements, gas and
total thermal energy ejected into the ICM by the cluster galaxies 
we integrate 
the contributions from the single galaxies over the cluster K-band luminosity function 
at each given cosmic time
for clusters of different mass (temperature). For the equations and related details
we direct the reader to Pipino et al. (2002).

In figure 2 we show the Fe abundance in the ICM predicted by our models 
compared to the observed one by White (2000) as functions of cluster temperature (left panel).
\bf Our best model (with Salpeter IMF and  evolution 
of Spirals) is shown by the long dashed line. \rm  
 \begin{figure}
           \includegraphics[width=6cm,height=6.5cm]{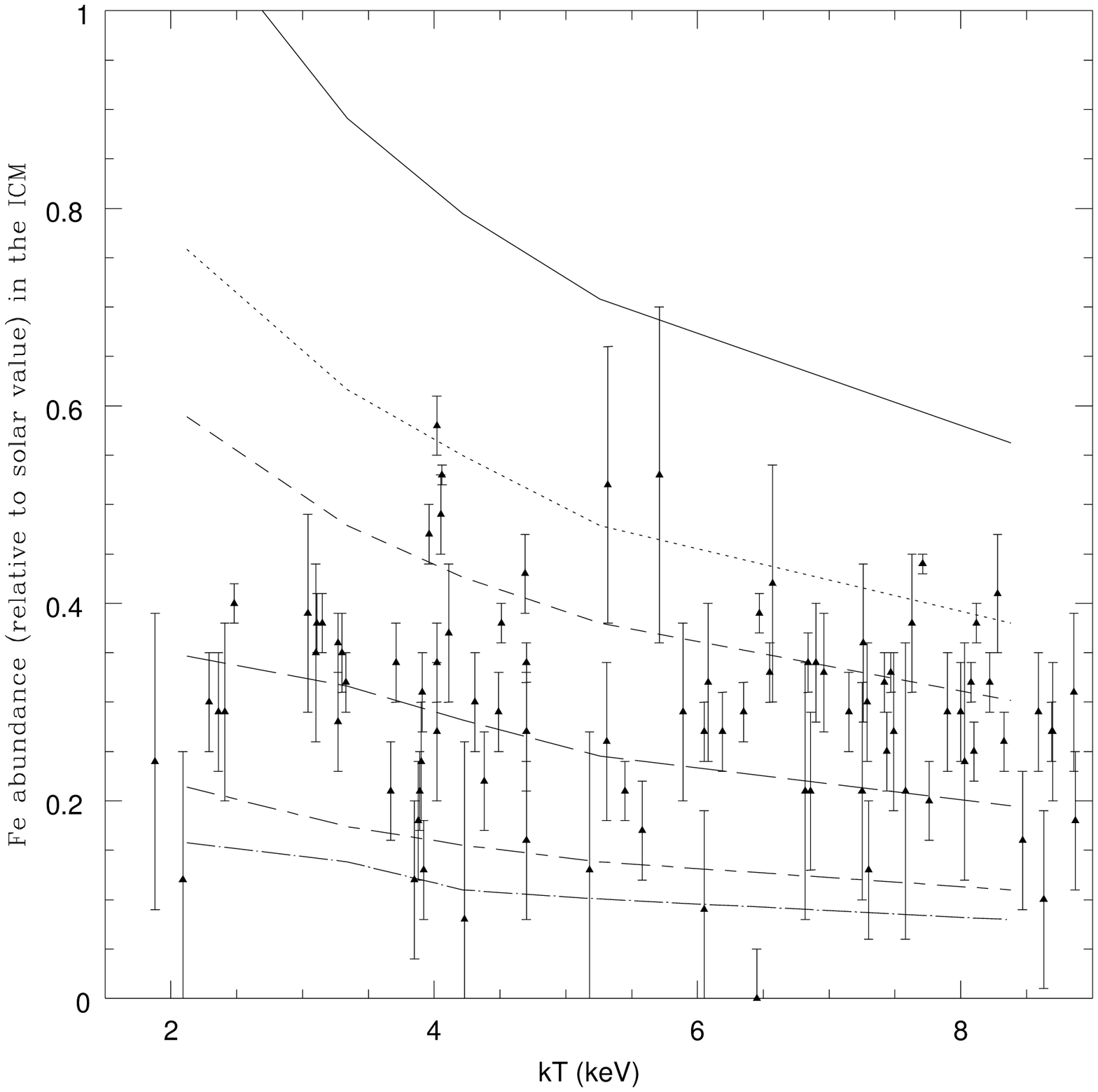}
           \includegraphics[width=6cm,height=6.5cm]{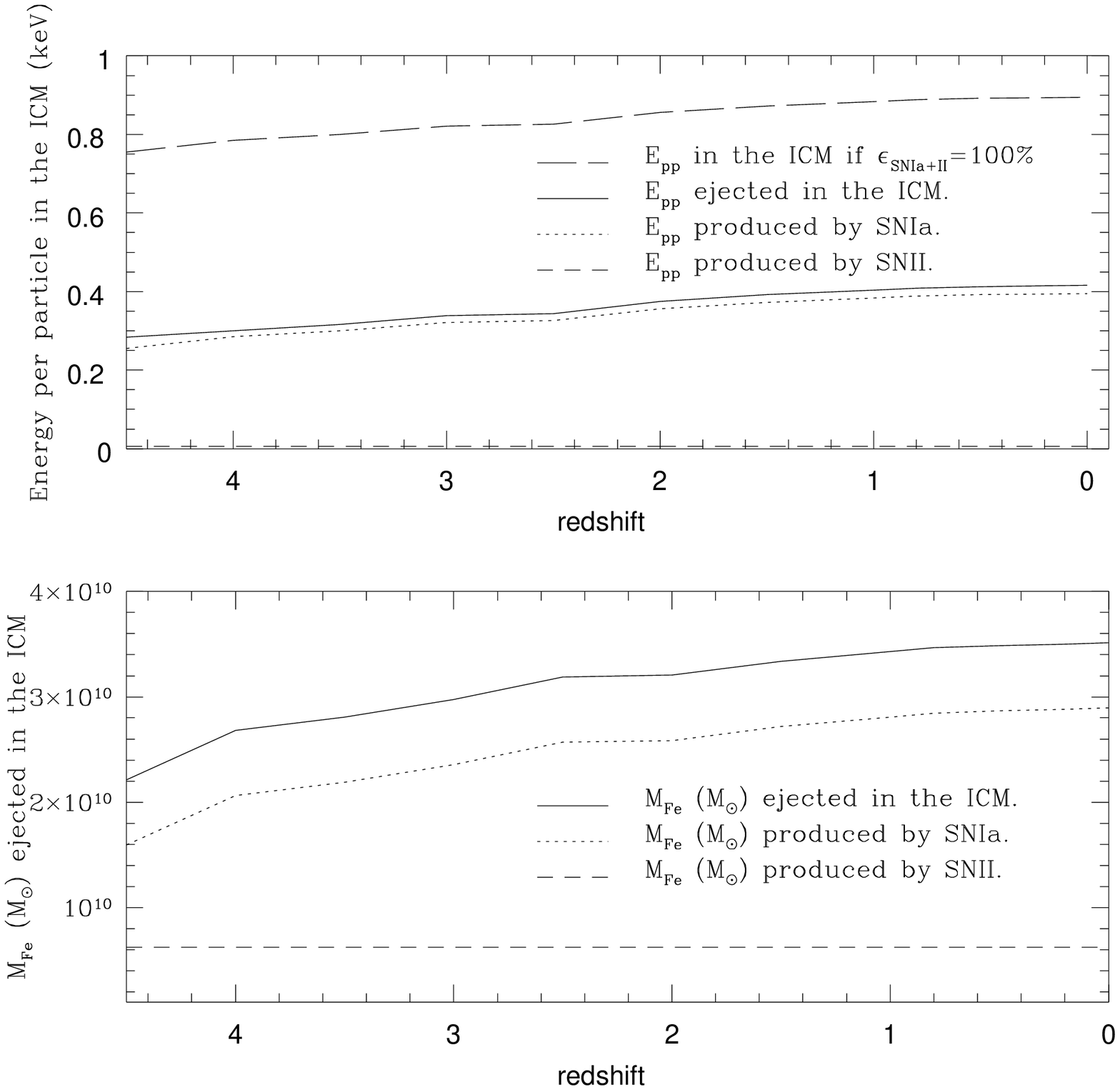}
	   \caption{\tiny \it Left: \rm Model predictions for the Fe abundance as a function of the temperature of the ICM
compared to the data by White (2000). \it Right: \rm predicted evolution in time of the Fe mass and the energy ejected by ellipticals into
the ICM.}
           \end{figure}
We have also computed the amount of metals which remains locked up 
inside stars. We found that the Fe in the ICM should be 5 times more than the Fe in stars,
whereas O and the global metal content should be roughly the same inside 
stars and in the ICM.

The best model can also reproduce the observed  [$\alpha$/Fe] ratios
      in the ICM. In particular, there is good agreement between the 
predicted ($\sim \rm solar$) and observed  ($\sim 0.1\pm 0.1 \rm dex$) [Si/Fe] ratios, 
whereas the predicted [O/Fe] ratios 
are lower ($\sim -0.3 \rm dex$) than the solar value, as indicated by some observations
(e.g. Ishimaru \& Arimoto 1997). 
However, the O measurements in the ICM 
are still very sparse and uncertain and more data are required to assess 
this point (but see Gastaldello \& Molendi 2002).

Models with the choice \bf b) \rm for the stellar yields from massive stars predict
the same amount of Fe and energy in the ICM, 
whereas the differences in the predictions for [$\alpha$/Fe] ratios are 
less than $\sim 0.2 \rm dex$.
             
On the right-hand side of figure 2 we show the predicted evolution of the energy per particle in the ICM,
$E_{pp}$ (upper panel) 
and the iron mass in the ICM (lower panel)
as a function of redshift by ``the best model''.
Little evolution is found both for the abundances and for the heating
energy  from z=0 up to z=1, in agreement with observations (Matsumoto et al. 2000). 
     
         \section{Conclusions}
While type II SNe dominate the 
chemical evolution of the ellipticals (creating $\alpha$ enhancement) 
, type Ia SNe play a fundamental role in
      providing energy ($\sim 80-95 \% $) and Fe ($\sim 45-80 \% $) 
      into the ICM.

Comparing masses and energy ejected by galaxies formed at $z=8$
with those from galaxies formed at $z\sim 3$ leads to the conclusion 
that the bulk of the ejection 
has to take place at $3 < z < 5 $.

The best model, which reproduces the Fe abundances in the ICM, 
can provide 0.20-0.35 keV of energy  per ICM particle,
depending on the cluster richness. 

           
\end{article}
\end{document}